\documentclass[12pt,preprint]{aastex}

\newcommand{\text}[1]{\rm #1}

\slugcomment{}

\shorttitle{Inelastic Neutrino-Helium Scatterings
 and Shock Instability in Core-Collapse Supernovae}
\shortauthors{Ohnishi et al.}

\begin{document}

\title{Inelastic Neutrino-Helium Scatterings
 and Standing Accretion Shock Instability in Core-Collapse Supernovae}

\author{Naofumi Ohnishi\altaffilmark{1}, Kei Kotake\altaffilmark{2},
and Shoichi Yamada\altaffilmark{3,4}}
\affil{$^1$Department of Aerospace Engineering, Tohoku University,
6-6-01 Aramaki-Aza-Aoba, Aoba-ku, Sendai, 980-8579, Japan}
\email{ohnishi@cfd.mech.tohoku.ac.jp}
\affil{$^2$ National Astronomical Observatory of Japan,
2-21-1, Osawa, Mitaka, Tokyo, 181-8588, Japan}
\email{kkotake@th.nao.ac.jp}
\affil{$^3$Science \& Engineering, Waseda University,
3-4-1 Okubo, Shinjuku, Tokyo, 169-8555, Japan}
\affil{$^4$Advanced Research Institute for Science and Engineering,
Waseda University,
3-4-1 Okubo, Shinjuku, Tokyo, 169-8555, Japan}
\email{shoichi@heap.phys.waseda.ac.jp}

\begin{abstract}
We present the results of numerical experiments,
in which we have investigated the influence
of the inelastic neutrino-helium interactions
on the standing accretion shock instability supposed
to occur in the post-bounce supernova core.
The axisymmetric hydrodynamical simulations of accretion flows
through the standing accretion shock wave onto the protoneutron star
show that the interactions are relatively minor
and the linear growth of the shock instability is hardly affected.
The extra heating given by the inelastic reactions
becomes important for the shock revival after the instability enters
the non-linear regime,
but only when the neutrino luminosity is very close
to the critical value,
at which the shock would be revived without the interactions.
We have also studied the dependence of the results
on the initial amplitudes of perturbation
and the temperatures of mu and tau neutrinos.
\end{abstract}
\keywords{supernovae: general --- neutrinos --- hydrodynamics
--- instabilities}

\clearpage

\section{Introduction}

Most of the supernova modelers are currently concerned
with the multi-dimensional aspects of the dynamics,
pushed by the accumulating observational
evidences that the core-collapse supernovae are
generally aspherical \citep{wang96,wang01,wang02}.
Various mechanisms to produce the asymmetry have been considered so far.
Among them are the convection
(e.g., \citet{herant_94,burrows_95,jankamueller96}),
growth of asymmetry seed generated prior
to core-collapse \citep{burohey,fryerkick},
rotation and magnetic fields
(see \citet{yamasawa,kotakemhd,takiwaki,sawa,ard}
and \cite{kotake_rev} for collective references).

Recently the standing accretion shock instability (SASI)
is attracting interest of researchers.
The instability was originally studied for transonic accretion flows
to black holes (e.g., \citet{foglizzo01,foglizzo02}) and
was re-discovered by \citet{blondin_03}
in the context of core-collapse supernovae.
\citet{blondin_03} found in their 2D numerical simulations
of the spherically symmetric, isentropic, steady accretion flows,
that a standing shock wave are unstable to non-spherical perturbations
and that the perturbations grow up to the nonlinear regime
with the clear dominance of the $\ell=1$ mode
at first and $\ell=2$ mode later,
leading to the global deformation of the shock wave as has been observed.
Here $\ell$ stands for the azimuthal index of the Legendre polynomials.

The mechanism of the instability is still controversial.
\citet{blondin_05} took coolings into account
in a simple analytic way following \citet{HC} and claimed that
the repeated propagations of pressure fluctuations are
responsible for the instability.
On the other hand,
\citet{ohnishi_1} did 2D numerical experiments,
implementing more realistic heatings as well as coolings by neutrino
and found that the original idea by \citet{foglizzo01,foglizzo02}
that the non-radial instability is driven
by the cycle of the advection of entropy and velocity fluctuations
and propagation of pressure perturbations seems to be more appropriate.
Obviously detailed linear analyses of the instability
are needed~\citep{ga05,yy06b}.

It is important that such a low-$\ell$-mode deformation
of the shock wave has been also observed
in more realistic simulations \citep{scheck_04}
and that the asymmetric explosion following the instability
might reproduce various non-spherical features of SN1987A~\citep{kifo}. 
It should be also emphasized that the instability is helpful
for the shock revival just as the convection is~\citep{ohnishi_1}.
The problem is whether the instability enhances the neutrino heating
sufficiently to revive the stalled shock.
Recent numerical investigations by~\citet{ja05} seem
to say ``no'' to this question.
It is true that we have to wait for detailed 3D simulations
before drawing a conclusion,
but we had better continue to seek for some other processes
that would further facilitate the shock revival.
As such a potential boost,
we will focus in this paper on the interplay
between the inelastic neutrino-nucleus interactions and the SASI.

The potential importance of the inelastic neutrino-nucleus interactions
was first pointed out by \citet{haxton},
who paid attention to the heating and dissociations of nuclei
in the matter ahead of the shock wave via these reactions,
the so-called preheating.
Taking the reactions into account
in their 1D spherically symmetric numerical simulations,
\citet{bruenn_haxton} found that the effect of preheating
is quite minor mainly because $\nu_{\rm e}$ energies obtained
in the simulation was lower than the one assumed in \citet{haxton}.
They also discussed the possibility of shock revival
by enhanced heating of the postshock material
by inelastic $\nu_{\tau, \mu}-{}^4{\rm He}$ scatterings,
the idea similar to ours pursued in this paper.
They found that the reactions are not very important.
It should be emphasized, however,
that the conclusion will be sensitive to the background model
and they considered a single snapshot after the bounce.
Furthermore, in the spherically symmetric models,
most of the nuclei are photodisintegrated after passing
through the shock and the reactions will scarcely occur anyway.
The situation may be different in non-spherical cases.
Since the shock wave hovers at larger radius in general,
not all the nuclei are dissociated and the heating region will be wider.
In this paper,
we pay particular attention to the interplay
between the SASI and the inelastic reactions.

The plan of this paper is as follows.
We describe the numerical methods,
input physics and models in section \ref{s2}.
The main numerical results are shown in section \ref{s3}.
We conclude this paper with section \ref{s4}.

\section{Numerical Method, Input Microphysics and Models\label{s2}}

In this paper we study the effect
of the inelastic neutrino-helium interactions on the evolution
of accretion flows through a shock wave onto a protoneutron star,
in particular the growth of SASI. 
We assume the axisymmetry of the system and do 2D numerical simulations. 

The numerical methods employed in this paper
are essentially the same as those used
in our previous paper \citep{ohnishi_1}.
The following equations describe the compressible accretion flows
of matter attracted by the protoneutron star
and irradiated by neutrinos emitted from the neutrino sphere.
\begin{equation}
 \frac{d\rho}{dt} + \rho \nabla \cdot \mbox{\boldmath$v$} = 0,
\end{equation}
\begin{equation}
 \rho \frac{d \mbox{\boldmath$v$}}{dt} = - \nabla P - \rho\nabla\Phi,
\end{equation}
\begin{equation}
 \rho \frac{d}{dt}\displaystyle{\Bigl(\frac{e}{\rho}\Bigr)}
  = - P \nabla \cdot \mbox{\boldmath$v$} + Q_{\text{E}}
  + Q_{\rm inel},
  \label{eq:energy}
\end{equation}
\begin{equation}
 \frac{dY_{\rm e}}{dt} = Q_{\text{N}},
  \label{eq:ye_flow}
\end{equation}
\begin{equation}
 \Phi = - \frac{G M_{\rm in}}{r},
  \label{eq:domain_g}
\end{equation}
where $\rho$, $\mbox{\boldmath$v$}$, $e$, $P$, $Y_{\rm e}$,
and $\Phi$ are density, velocity, internal energy, pressure,
electron fraction, and gravitational potential, respectively.
We denote the Lagrangian derivative as $d/dt$ and $r$ is the radius.
The self-gravity of matter in the accretion flow
is ignored (see \citet{yy06a} for the effect).
The parameters of $Q_{\rm E}$ and $Q_{\rm N}$ are related
with the interactions of neutrinos
and free nucleons (see also \citet{ohnishi_1}).
$M_{\rm in}$ is the mass of the central object.

In addition to the standard heating
and cooling via neutrino absorptions and emissions by free nucleons,
here we consider the inelastic neutrino-helium interactions.
The heating rates denoted as $Q_{\rm inel}$ were estimated
by \citet{haxton} for the inelastic scatterings
on nuclei via neutral currents,
$\nu + (A,Z) \rightarrow \nu + (A,Z)^{*}$, as follows,
\begin{eqnarray}
 Q_{\text{inel}}
  = \frac{\rho X_{\text{A}}}{m_{\text{B}}}
  \frac{31.6\text{MeV}}{(r/10^{7}\text{cm})^{2}}
  &&\left[
     \frac{L_{\nu_{\text{e}}}}{10^{52}\text{ergs~s}^{-1}}
     \left(\frac{5\text{MeV}}{T_{\nu_{\text{e}}}}\right)
     \frac{A^{-1}\langle
     \sigma_{\nu_{\text{e}}}^{+}E_{\nu_{\text{e}}}
     +\sigma_{\nu_{\text{e}}}^{0}E_{\text{ex}}^{\text{A}}
     \rangle_{T_{\nu_{\text{e}}}}}
     {10^{-40}\text{cm}^2\text{MeV}} \right. \nonumber \\
 &&+\quad\frac{L_{\bar{\nu}_{\text{e}}}}{10^{52}\text{ergs~s}^{-1}}
  \left(\frac{5\text{MeV}}{T_{\bar{\nu}_{\text{e}}}}\right)
  \frac{A^{-1}\langle
  \sigma_{\bar{\nu}_{\text{e}}}^{-}E_{\nu_{\text{e}}}
  +\sigma_{\bar{\nu}_{\text{e}}}^{0}E_{\text{ex}}^{\text{A}}
  \rangle_{T_{\bar{\nu}_{\text{e}}}}}
  {10^{-40}\text{cm}^2\text{MeV}} \nonumber \\
 &&+\left.\quad
    \frac{L_{\nu_{\mu}}}{10^{52}\text{ergs~s}^{-1}}
    \left(\frac{10\text{MeV}}{T_{\nu_{\mu}}}\right)
    \frac{A^{-1}\langle
    \sigma_{\nu_{\mu}}^{0}E_{\text{ex}}^{\text{A}}
    +\sigma_{\bar{\nu}_{\mu}}^{0}E_{\text{ex}}^{\text{A}}
    \rangle_{T_{\nu_{\mu}}}}
    {10^{-40}\text{cm}^2\text{MeV}}
   \right],
 \label{eq:inelastic}
\end{eqnarray}
where $X_{\rm A}$ is the mass fraction of the nucleus
and $m_{\rm B}$ is the atomic mass unit.
$L_{\nu}$ and $T_{\nu}$ in the square brackets are
the neutrino luminosity and temperature, respectively,
and $A$ is the mass number of the nucleus.
The last term denotes the sum of the contributions
from mu and tau neutrinos.
The cross section for each neutral-current is evaluated
by the following fitting formula,
\begin{equation}
 A^{-1}\langle
  \sigma_{\nu}^{0}E_{\text{ex}}^{\text{A}}
  +\sigma_{\bar{\nu}}^{0}E_{\text{ex}}^{\text{A}}
  \rangle_{T_{\nu}}
  = \alpha\left[ \frac{T_{\nu} - T_{0}}{10\text{MeV}} \right]^{\beta},
  \label{eq:fitting}
\end{equation}
where $\alpha$, $\beta$, and $T_{0}$ are given
in Table~I of \citet{haxton}.
Since we are concerned with the reactions with $^{4}{\rm He}$,
the only nucleus that is abundant in the post shock matter,
these parameters are chosen to be
$\alpha=1.24\times 10^{-40}$~MeV~cm$^{2}$, $\beta=3.82$,
and $T_{0}=2.54$~MeV.
In the first and second terms on the right hand side
of Eq.~(\ref{eq:inelastic}),
the contributions from the charged current reactions,
$\sigma_{\nu}^{+}$ and $\sigma_{\nu}^{-}$,
are also taken into account according to Table~II of \citet{haxton}.
We ignore the variations of the electron fraction by these reactions,
since they are minor and give no qualitative difference to the dynamics.
Considering the uncertainties inherent to the theoretical estimation
of the reaction rates,
we multiply rather arbitrarily the rates obtained above
and discuss the dependence of the outcomes on this factor.

The numerical code employed
in this paper is based on the ZEUS-2D \citep{stone},
which is an Eulerian code based on the finite-difference method
with an artificial viscosity of von Neumann and Richtmyer type.
We have made several major changes to the base code
to include appropriate microphysics.
For example,
we have added the equation for electron fraction (Eq.~(\ref{eq:ye_flow})),
which is solved in the operator-splitting fashion.
We have also incorporated the tabulated realistic equation of state (EOS) 
based on the relativistic mean field theory \citep{shen98}
instead of the ideal gas EOS assumed in the original code.
The reason why only $^{4}{\rm He}$ is considered in this paper is
that the abundance of other nuclei is negligibly small
in the post-shock matter.
The mass fraction of $^{4}{\rm He}$ is obtained from the EOS.

Spherical coordinates are used.
No equatorial symmetry is assumed and the computation domain covers
the whole meridian section with 60 angular mesh points,
except for a model in which we have adopted 120 angular mesh points.
Since the latter model did not produce any significant difference
from other models, 
we will report in the following the results obtained
from the models with 60 angular mesh points.
We use 300 radial mesh points to cover
$r_{\rm in} \leq r \leq r_{\rm out} = {2000}$~km,
where $r_{\rm in}$ is the inner boundary and
chosen to be roughly the radius of neutrino sphere.

The initial conditions are prepared in the same manner
as in \citet{ohnishi_1}.
The steady state solutions obtained by \citet{yamasaki}
for a fixed density at the inner boundary,
$\rho_{\rm in} = 10^{11}$~g~cm$^{-3}$, are utilized.
In so doing, $Q_{\rm inel}$ is not taken into account.
Hence the initial state is not completely steady
when the inelastic interactions are considered
and this slight inconsistency can be regarded
as an additional radial perturbation.
As shown shortly, however,
the effect is small and limited to a very narrow region,
and matters little to the analysis of the following dynamics.
To induce the non-spherical instability,
we have added $\ell = 1$ velocity perturbations
to the initial state mentioned above.

All the numerical models are summarized in Table~\ref{tab:model}.
The mass accretion rate and the mass of protoneutron star are fixed
to be $\dot{M} = 1~M_{\odot}$~s$^{-1}$ and $M_{\rm in} = 1.4~M_{\odot}$,
respectively.
The temperatures of electron-type neutrinos are also constant
and set to be $T_{\nu_{\text{e}}} = 4$~MeV
and $T_{\bar{\nu}_{\text{e}}} = 5$~MeV,
which are the typical values in the post-bounce phase.
For most of the models,
the temperature of mu and tau neutrinos is chosen
to be $T_{\nu_{\mu}} = 10$~MeV,
but we also vary it to investigate the dependence of the dynamics
on this parameter.
Note that the reaction rates are very sensitive to the incident energy
of neutrino (see Eq.~(\ref{eq:fitting})).
The neutrino luminosity is also varied in this study.
In the reference model,
the luminosity of electron-type neutrino $L_{\nu_{\rm e}}$
and anti-neutrino $L_{\bar{\nu}_{\rm e}}$ are
set to be $5.9\times 10^{52}$~ergs~s$^{-1}$.
It is noted that this value is very close to the threshold,
$L_{\nu_{\rm e}, \bar{\nu}_{\rm e}} = 6.0\times 10^{52}$~ergs~s$^{-1}$,
at which a SASI-triggered shock revival occurs
without inelastic interactions as described in \citet{ohnishi_1}.
The luminosity of mu and tau neutrinos is set to be half
that of electron-type neutrinos according to the results
obtained by detailed simulations (e.g., \citet{lieb01}).

\section{Results \label{s3}}

\begin{figure}
\epsscale{0.8}
\plotone{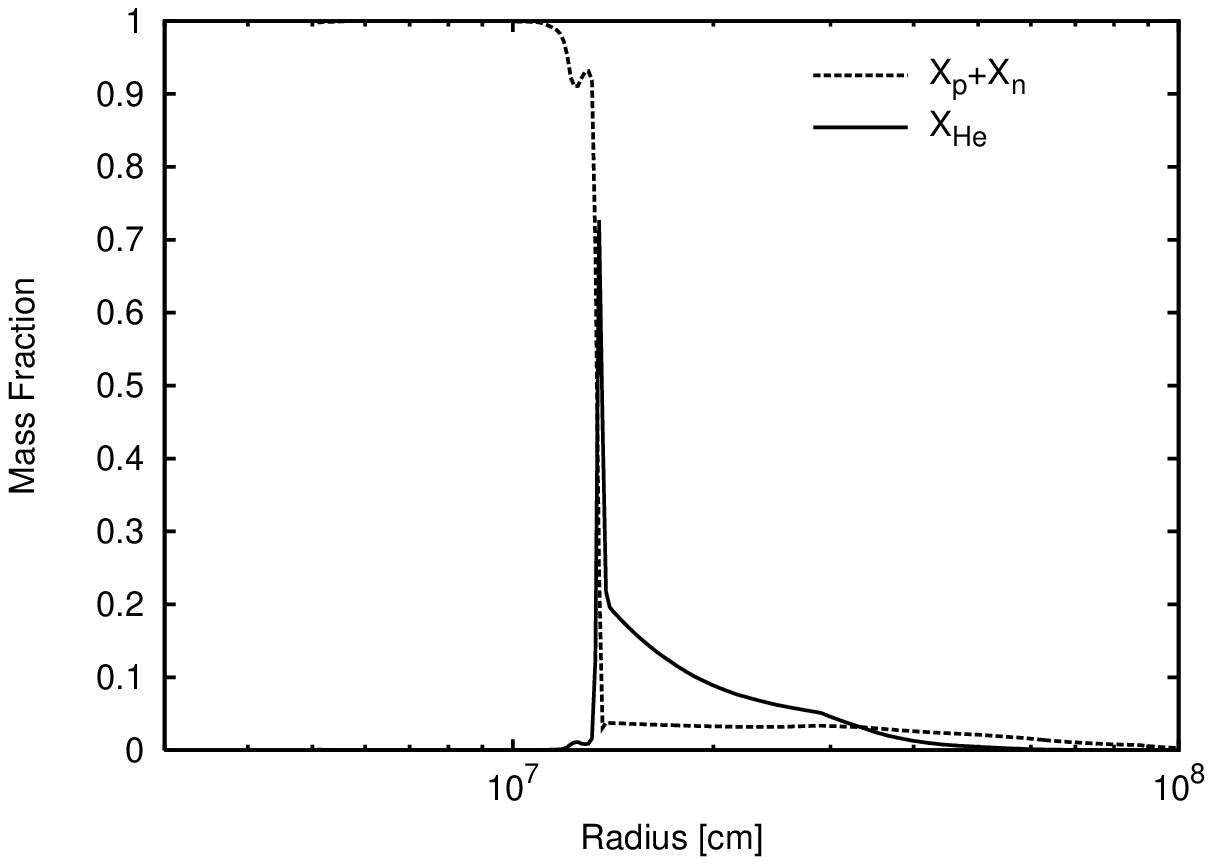}\\
\plotone{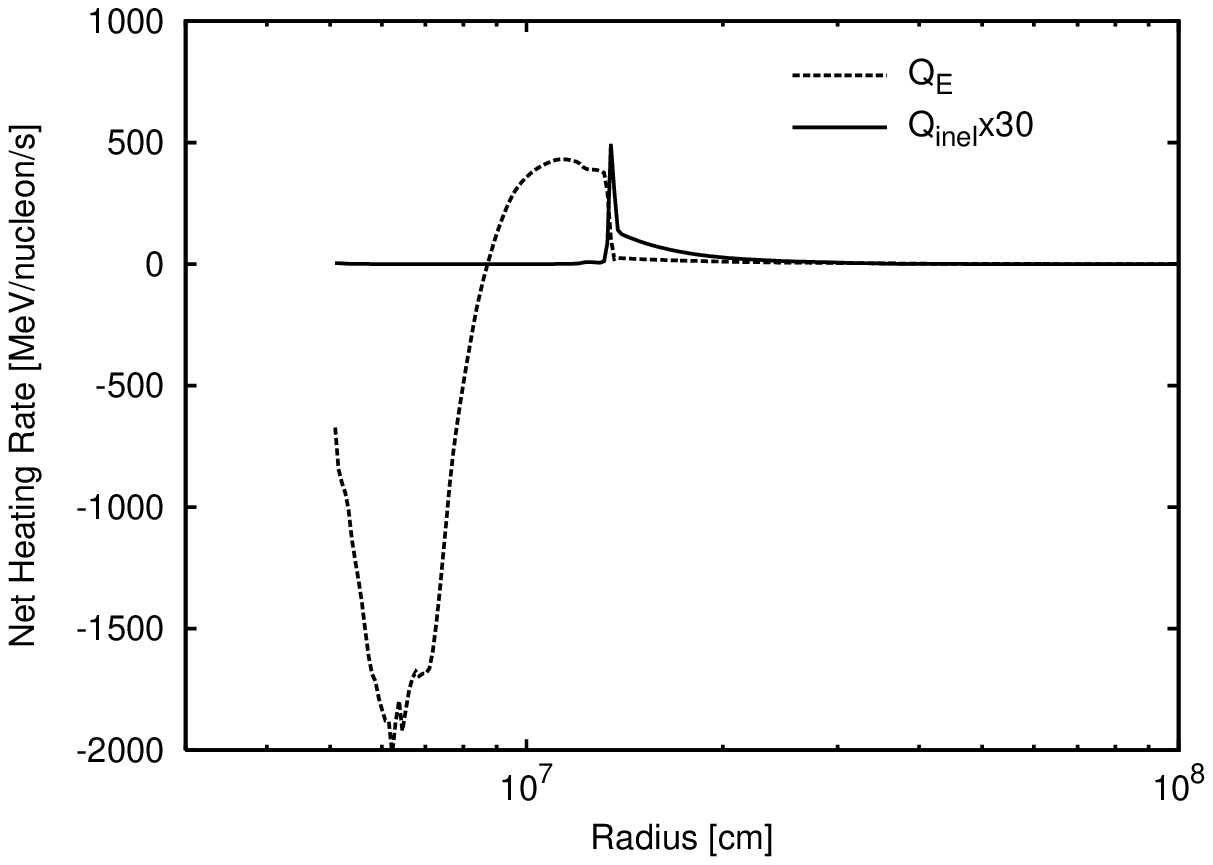}
\caption{%
The mass fractions of helium and nucleons (upper panel)
and the heating rates (lower panel)
at the initial time for the reference model L59I0.
The solid and dashed lines in the upper panel denote
the mass fractions of helium and nucleons, respectively.
The lower panel represents the net heating rates
by absorptions and emissions on nucleons (solid line)
and the inelastic interactions with helium (dashed line).
Note that the latter rate is multiplied by a factor of 30.
}
\label{fig:profiles}
\end{figure}

\begin{figure}
\epsscale{1.0}
\plottwo{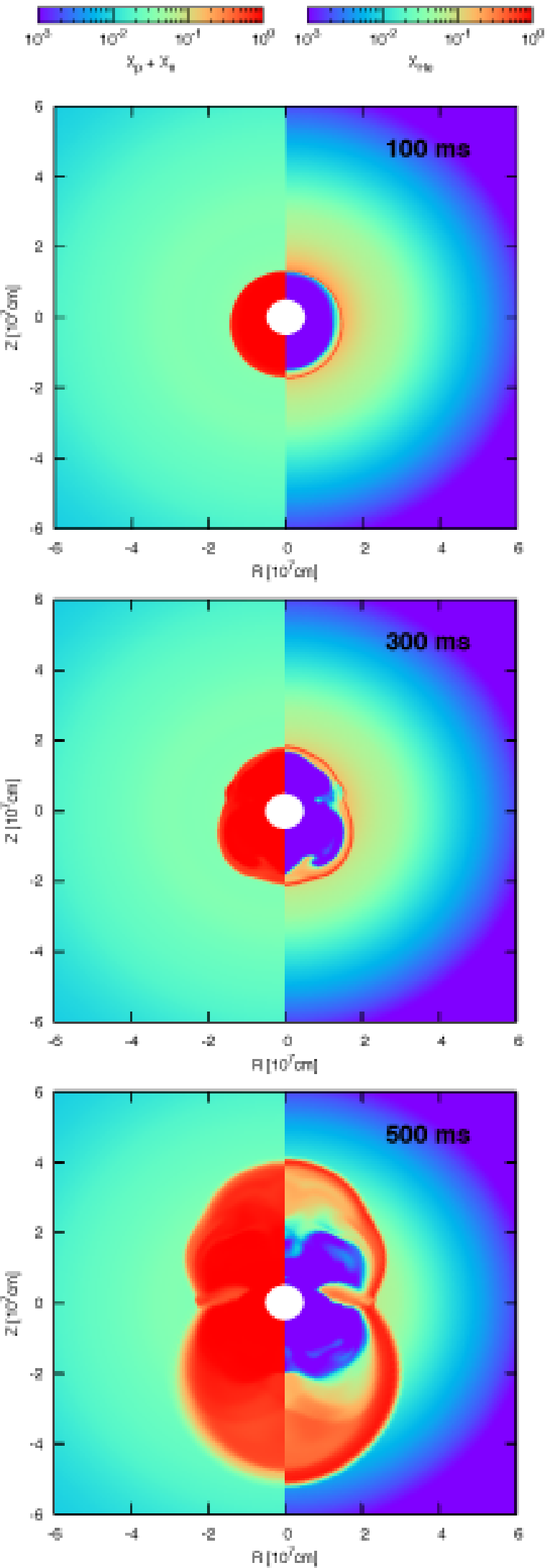}{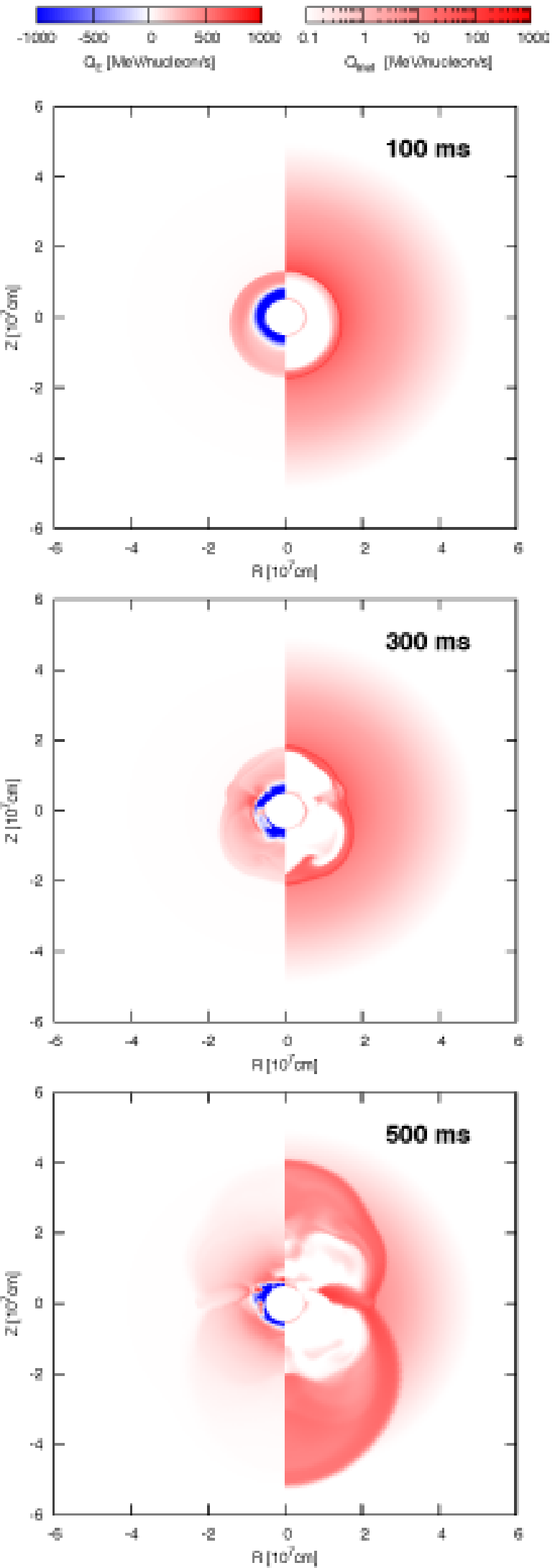}
\caption{%
The contours in the meridian section of the mass fractions
(left panels) and net heating rates (right panels) for model of L59I10.
The left and right halves of each left panel represent
the fractions of nucleons and helium, respectively.
The heating rates are plotted for the neutrino-nucleon
reactions (left half) and the inelastic interactions
of neutrino with helium (right half).
Note the latter is in the logarithmic scale.
}
\label{fig:contour}
\end{figure}

Figure~\ref{fig:profiles} shows the mass fractions of proton,
neutron and helium (upper panel)
and the profiles of $Q_{\rm E}$ and $Q_{\rm inel}$ (lower panel)
at the initial time for the reference model L59I0,
where $L_{\nu_{\rm e}}$ is set to be $5.9\times 10^{52}$~ergs~s$^{-1}$
and $T_{\nu_{\mu}} = 10$~MeV.
The helium abundance is small except for a narrow region
inside the shock wave.
All the nuclei are completely dissociated to nucleons
after passing through the shock wave
because the standing shock is located deep
inside the gravitational potential-well
in spherically symmetric accretions and, as a result,
the post-shock temperature becomes too high for nuclei to survive.
There is also a small population of helium ahead of the shock
owing to the partial decomposition of nuclei by adiabatic compressions.
This small abundance is the main reason
why most of the detailed numerical simulations have not incorporated
the reactions of neutrino with helium so far.
The heating by the inelastic interactions is appreciable
only inside the shock wave accordingly.
Note also that the value of $Q_{\rm inel}$ is multiplied
by a factor of 30 in the figure.
It is thus expected that the inelastic reactions will not affect
the dynamics at least in the initial phase.
This may not be the case for later phases, however.
After the non-spherical instability grows,
the shock radius becomes larger in general and, as a result,
the helium abundance will be increased in a wider region.
Moreover, most of these helium will be populated
in the so-called heating region (see Fig.~\ref{fig:contour}).

\begin{figure}
\epsscale{1.0}
\plotone{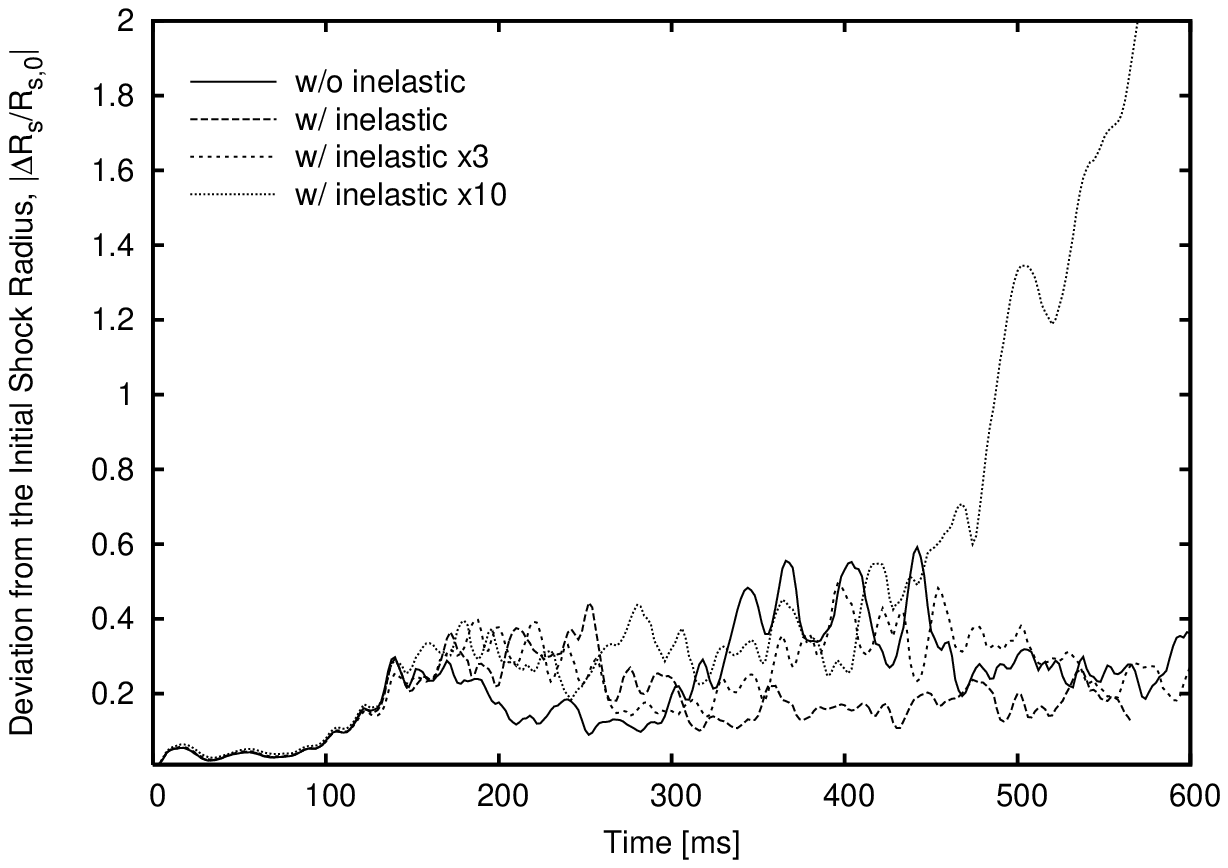}
\caption{%
The temporal evolutions of the angle-averaged shock radius 
for the models with $L_{\nu_{\rm e}} = 5.9\times 10^{52}$ ergs~s$^{-1}$.
The relative deviations from the initial value are plotted
for models of L59I0, L59I1, L59I3, and L59I10.
}
\label{fig:shock_radius_l1_001}
\end{figure}

We first summarize the basic feature of the temporal evolution
of the reference model L59I0
after 1\% of the $\ell = 1$ single-mode velocity perturbation is added. 
The exponential growth of the perturbation is observed at first
and the shock surface is deformed by the increasing amplitude
of the non-radial mode.
When the non-linear regime is reached,
the shock begins to oscillate with a large amplitude.
As shown in Fig.~\ref{fig:shock_radius_l1_001},
where the time evolution of the angle-averaged shock radius is presented,
the oscillation becomes quasi-steady by $\sim$150~ms.
Note that the shock radius in this phase is larger than
the initial value as pointed by \citet{ohnishi_1}.
We have found no shock revival for this model.
In fact, as mentioned already,
the shock revival is found only
for $L_{\nu_{\rm e}}\geq6\times 10^{52}$~ergs~s$^{-1}$
if the inelastic interactions are not taken into account.
In the last column of Table~\ref{tab:model},
we summarize for each model if the shock revival is found
by $\sim$500~ms after the onset of computation.
It should be noted that the shock revival, if observed in our models,
does not guarantee the explosion in more realistic settings,
since the neutrino luminosity will not be constant in time
as assumed in our models and will decline in reality.
Hence our criterion for the shock revival should be regarded
as a minimum requirement for explosion.

\begin{figure}
\epsscale{1.0}
\plotone{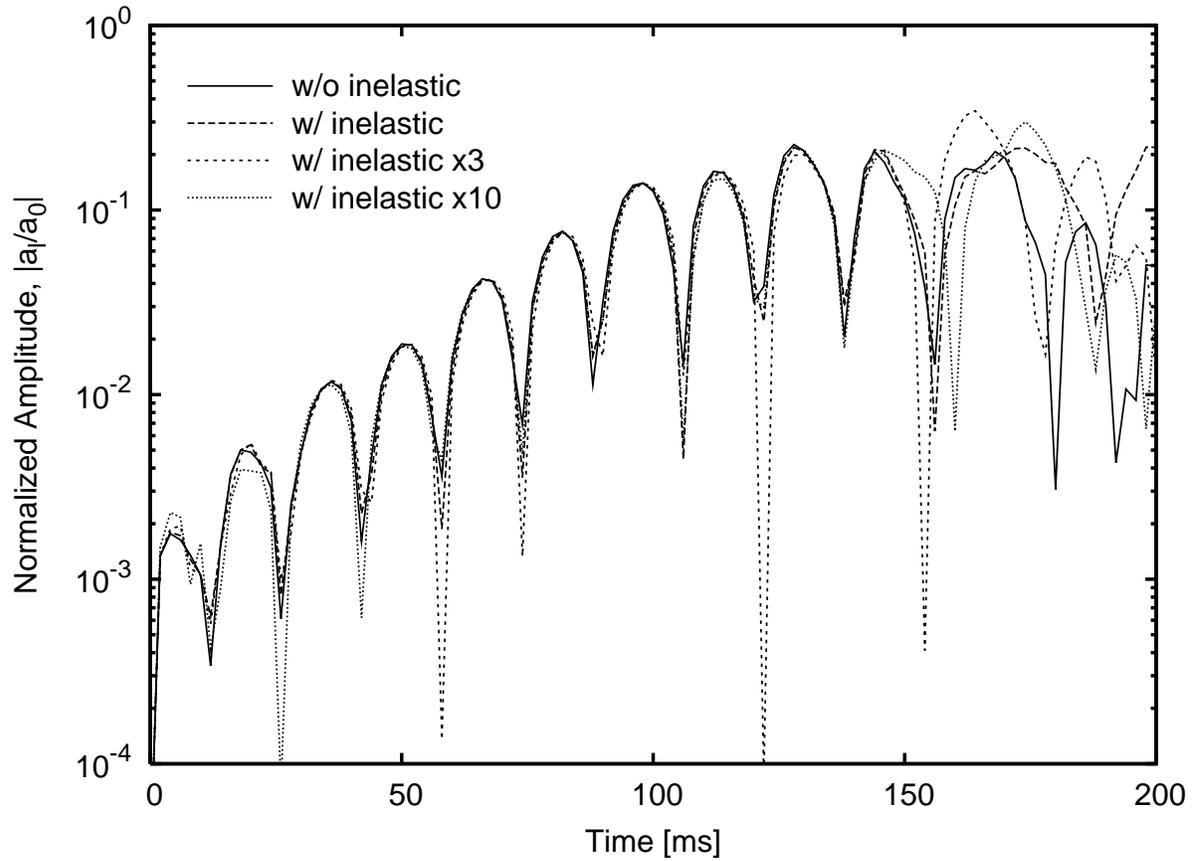}
\caption{%
The temporal evolutions of the normalized amplitudes
of the $\ell = 1$ mode in the spherical harmonic decompositions
for models L59I0, L59I1, L59I3 and L59I10.
See the text for details.
}
\label{fig:l1amplitude}
\end{figure}

Now we proceed to consider the effect of the inelastic interactions
of neutrinos with helium.
The time evolutions of shock radius for models L59I1, L59I3, and L59I10
are presented in Fig.~\ref{fig:shock_radius_l1_001}
together with that for the reference model L59I0.
These models have the same neutrino luminosity as the reference model
and are given the initial velocity perturbation of 1\%.
The difference is the assumed cross sections for the inelastic reactions.
As mentioned earlier,
considering the uncertainties
that the theoretical estimation inherently has,
we multiply the nominal values of the cross sections given
by Eqs.~(\ref{eq:inelastic}) and (\ref{eq:fitting}),
by the factors given in Table~\ref{tab:model}.
Except for model L59I10,
the shock oscillations accompanied by the growth of SASI are settled
to quasi-steady states by $\sim$150~ms just as in the reference model.
The final shock radii are not very different from each other
among these no-revival models
and are larger than that of the initial condition.
Model L59I10, whose $Q_{\rm inel}$ is multiplied by a factor of 10,
gives a shock revival after a rather long time, $\sim$450~ms.
As seen in Fig.~\ref{fig:shock_radius_l1_001},
the evolution in the early phase is essentially the same
as for other models,
as expected from the helium abundance in the initial condition.
This is also seen in the growth rates of the $\ell = 1$ mode
presented in Fig.~\ref{fig:l1amplitude}.
Here we decompose the deformation of the shock surface
into the spherical harmonic components;
\begin{equation}
 R_{\rm s}(\theta) = \sum_{\ell=0}^{\infty}a_{\ell}
  \sqrt{\frac{2\ell+1}{4\pi}}P_{\ell}(\cos\theta).
\end{equation}
Since the system is axisymmetric, only $m=0$ harmonics,
nothing but Legendre polynomials, show up.
The coefficients, $a_{\ell}$, can be calculated
by the orthogonality of the Legendre polynomials;
\begin{equation}
 a_{\ell} = \frac{2\ell+1}{2}\int_{-1}^{1}R_{\rm s}(\theta)
  P_{\ell}(\cos\theta)d\cos\theta.
\end{equation}
The position of the shock surface, $R_{\rm s}(\theta)$,
is determined as the iso-entropic surface of $s=5$.
No essential difference can be seen both in the linear phase
lasting for $\sim$150~ms and the early non-linear phase.
Therefore, the additional heating from the inelastic interactions
does not play an important role in the growth of SASI.

Figure~\ref{fig:contour} shows in the meridian section
the contours of the mass fractions of nucleons and helium
and the neutrino-heating rates for model L59I10.
Note that the heating rates for the inelastic reactions
(the right half of the right panels of Fig.~\ref{fig:contour})
are plotted in the logarithmic scale
whereas those for the others (the left half) are plotted
in the linear scale.
At 100~ms when the perturbation is still growing in the linear regime,
the mass fraction of helium is not so large in most of the region. 
One can see some minor heatings via the inelastic interactions
both inside and ahead of the shock wave,
the latter of which is the preheating considered
by \citet{haxton} and \citet{bruenn_haxton}.
At 300~ms, however,
the shock front wobbles and is deformed substantially by the SASI
in the non-linear regime,
and a part of the shock reaches larger radii from time to time
and the region behind the portion of the shock front contains
non-negligible fraction of helium.
This is simply because the temperature becomes lower there
and nuclei are not completely dissociated.
As a result of this increased helium population,
the neutrino heating is also enhanced, 
which then pushes the shock wave further outwards
and increases the volume, in which the helium is abundant.
This positive feedback finally leads to the shock revival
around 500~ms in this model.
One can see at this time that most of the region
behind the shock contains a large fraction of helium.

\begin{figure}
\epsscale{1.0}
\plotone{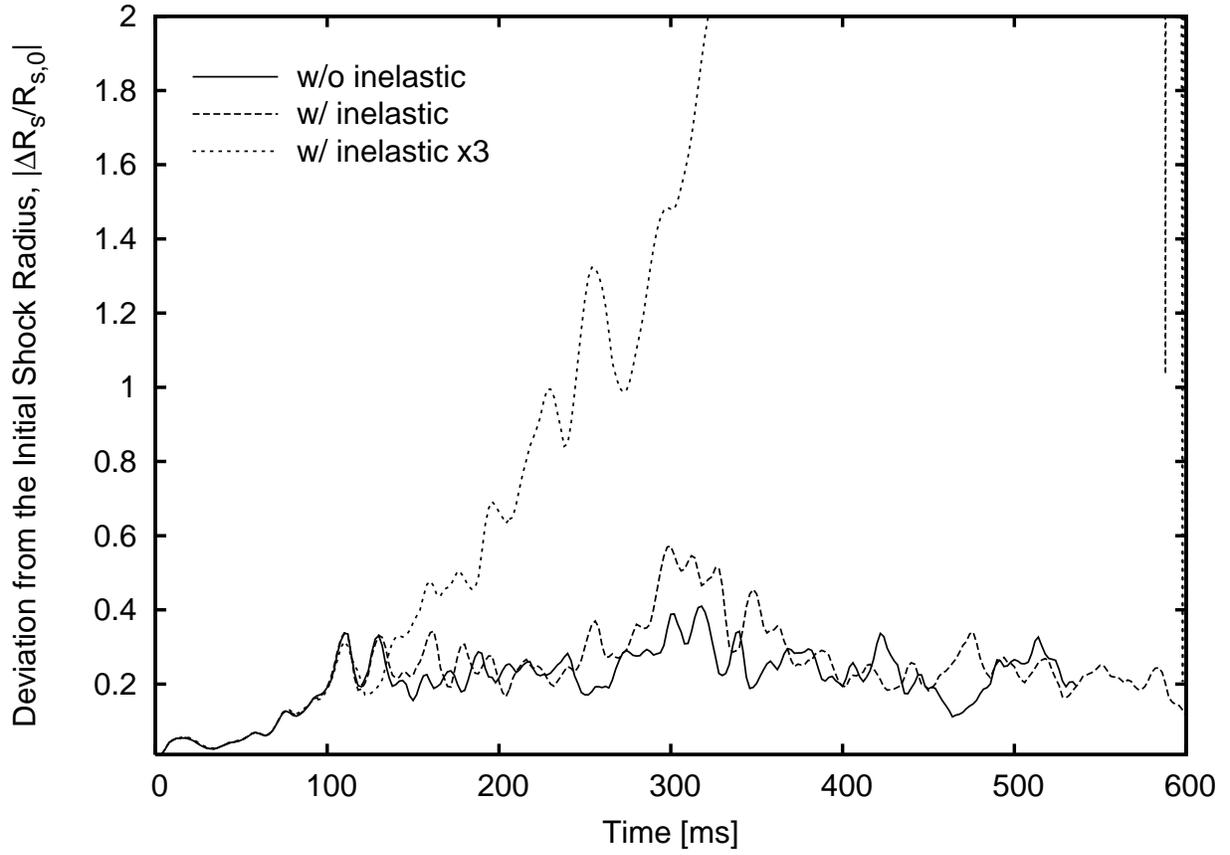}
\caption{%
The temporal evolutions of the angle-averaged shock radius
for the models with 5\% of initial velocity perturbation
($L_{\nu_{\rm e}} = 5.9\times 10^{52}$ ergs~s$^{-1}$).
The relative deviations from the initial value are plotted
for models L59I0d5, L59I1d5, and L59I3d5.
}
\label{fig:shock_radius_l1_005}
\end{figure}

We point out here that the shock revival is rather sensitive
to the initial amplitude of perturbation as demonstrated
in Fig.~\ref{fig:shock_radius_l1_005},
where we show the evolutions of the angle-averaged shock radius
for models L59I0d5, L59I1d5, and L59I3d5.
In these models we have imposed 5\%, instead of 1\%,
of the initial velocity perturbation.
We can observe that the perturbation grows more rapidly in these models.
The non-linear regime is reached in $\sim$100~ms.
More importantly, the shock revival occurs even for model L59I3d5
with the cross sections of the inelastic reactions multiplied
by a factor of 3 rather than 10,
the value required for the initial perturbation of 1\%.
Note, however, that the shock revival is achieved
without inelastic reactions if we add 10\%
of velocity perturbation initially (model L59I0d10).

\begin{figure}
\epsscale{1.0}
\plotone{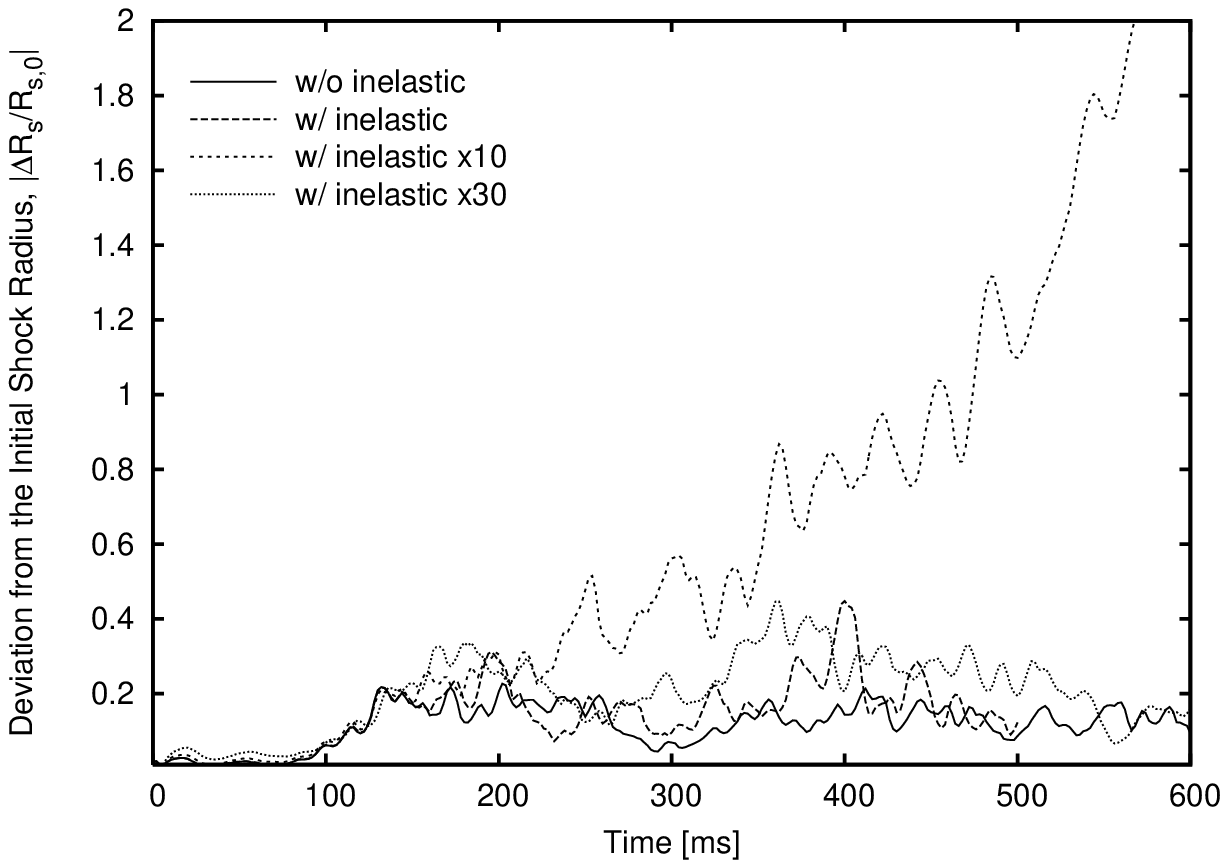}
\caption{%
The temporal evolutions of the angle-averaged shock radius
for the models with $L_{\nu_{\rm e}} = 5.8\times 10^{52}$ ergs~s$^{-1}$.
The relative deviations from the initial value are plotted
for models L58I0, L58I1, L58I10, and L58I30.
}
\label{fig:shock_radius_l1_001_Ln58}
\end{figure}

It should be also mentioned that the inelastic interactions lose
its importance very quickly as the neutrino luminosity is decreased.
As shown in Table~\ref{tab:model},
the models with $L_{\nu_{\rm e}}=5.5\times 10^{52}$~ergs~s$^{-1}$
have not led to the shock revival
even if we have multiplied the reaction rates by a factor of 30
(in fact, we have found a factor of 300 is required at least
in this model).
An interesting thing is that for models
with $L_{\nu_{\rm e}}=5.8\times 10^{52}$~ergs~s$^{-1}$,
the shock revival has been found for model L58I10
that has a multiplicative factor of 10
whereas the model with the factor of 30 (model L58I30) has not produced
a shock revival (see Fig.~\ref{fig:shock_radius_l1_001_Ln58}).
As shown in Fig.~\ref{fig:profiles},
the mass fraction of helium just ahead of the shock is $\sim$0.2 initially.
It seems that the preheating caused by these helium cannot be ignored
in this case and, in fact,
it tends to suppress the shock oscillations in model L58I30.
Although we have ignored the preheating by other nuclei in this paper,
they should be taken into account when the critical luminosity
is evaluated quantitatively.

Finally we discuss the dependence of the results
on the neutrino temperature.
As can be understood from Eq.~(\ref{eq:fitting}),
the inelastic scattering rates are very sensitive to the energy
of the incident neutrino ($\propto T_{\nu}^{3.8}$).
This naturally leads us to the question what will happen
if the energy spectra of neutrinos are harder than commonly assumed.
Since higher-energy neutrinos are more important,
here we modify only $T_{\nu_{\mu}}$.
The results are given in Table~\ref{tab:model} as model L59T-series.
Unfortunately, the results are not so sensitive
to the neutrino temperature as the cross sections themselves.
It is found that for $L_{\nu_{\rm e}} = 5.9\times 10^{52}$~ergs~s$^{-1}$
the extra heating by the inelastic interactions is large
enough to revive the stalled shock
only when the temperature of mu and tau neutrinos is higher
than $T_{\nu_{\mu}} = 25$~MeV.
This is much larger than the canonical value $\lesssim$10~MeV
and is highly unlikely to be obtained in the supernova core
(e.g., \citet{lieb01}).

\section{Summary and Discussion \label{s4}}

We have investigated the possible effects of the inelastic interactions
of neutrino with helium on the shock revival
in the post-bounce supernova core.
In particular, we have paid attention to their influence on the SASI,
one of the major causes for the asymmetry of dynamics
and a possible trigger of explosion.
For the spherically symmetric models,
\citet{bruenn_haxton} found that both the preheating of matter
ahead of the shock and heating of matter behind the shock
via these reactions are quite minor.
In fact, most of the nuclei are photodisintegrated
after passing through the shock
and the reactions will scarcely occur anyway
in the spherically symmetric models.
The situation may be different in non-spherical cases,
where the shock wave hovers at larger radius in general
and not all the nuclei are dissociated and the heating region is wider. 
In fact, these reactions have never been explored
in the multi-dimensional context so far.
We have done 2D numerical experiments on the post-bounce accretion flows
through the stalled shock wave onto the protoneutron star,
systematically changing the luminosity and temperature of neutrino
and the initial amplitude of perturbation as well as the reaction rates.

We have found that the incorporation of the inelastic interactions
has essentially no influence on the growth of the SASI,
since very little helium is existing in the post-shock matter initially.
However, these reactions become appreciable later
when the SASI enters the non-linear regime
and the shock oscillates with large amplitudes.
It has been shown that the extra heating by these interactions is
helpful for the shock revival in principle.
This is, however,
true in practice only when the shock revival is not obtained
with a slight margin without the interactions.
In fact, we have observed that a small ($\sim$10\%) reduction
of the neutrino luminosity makes the interactions entirely negligible.
Hence it is understandable that larger initial amplitudes
of perturbation make the interactions more important for shock revival.
It is, however, mentioned that even if the luminosity is very close
to the critical value,
the cross sections estimated by \citet{haxton} seem to be too small.
Although it is not easy to evaluate the uncertainties 
of the theoretical prediction,
recent new calculations by \citet{gazit04} may be used as a guide.
They predicted a bit larger $\beta$ than \citet{haxton}.
However, the enhancement of the heating rate for our reference model
is only 15\%,
much too small for the interactions to have some influence
on the shock revival.
The fact that the dynamics is rather insensitive
to the neutrino temperature is not encouraging, either.
Hence we conclude that the inelastic interactions of neutrino
with helium will be important only in determining
the shock-revival-point precisely.
It is, however, noted finally
that the inelastic neutrino-nuclei interactions should be incorporated
in the realistic simulations,
since, as we have seen, the preheating may suppress
the non-spherical oscillations of the shock wave.

\acknowledgements{
We are thankful to T. Yamasaki and M. Watanabe
for providing us with the information on their models.
K.~K. expresses thanks to K. Sato for continuing encouragement.
The numerical calculations were partially done
on the supercomputers at RIKEN and KEK
(KEK supercomputer projects 02-87 and 03-92).
This work was supported in part by Grants-in-Aid for Scientific Research
from the Ministry of Education, Science and Culture of Japan
(Nos. S14102004, 14079202, and 14740166),
and a Grant-in-Aid for the 21st century COE program
``Holistic Research and Education Center for Physics of
Self-organizing Systems.''}

\clearpage

\begin{deluxetable}{cccccccc}
\tabletypesize{\scriptsize}
\tablecaption{Model Parameters\label{tab:model}}
\tablewidth{0pt}
\tablehead{
\colhead{Model}
 & \colhead{$L_{\nu_{\rm e}}$ ($10^{52}$ ergs~s$^{-1}$)}
 & \colhead{$Q_{\rm inel}$ (Eq.~(\ref{eq:inelastic}))}
 & \colhead{$\delta v_{r}/v_{r}^{1D}$ (\%)}
 & \colhead{$T_{\nu_{\mu, \tau}}$ (MeV)}
 & \colhead{Shock Revival}}
\startdata
 L59I0    & 5.9 & --          & 1  & 10 & X\\
 L59I1    & 5.9 & $\times 1$  & 1  & 10 & X\\
 L59I3    & 5.9 & $\times 3$  & 1  & 10 & X\\
 L59I10   & 5.9 & $\times 10$ & 1  & 10 & $\bigcirc$\\
 L59I30   & 5.9 & $\times 30$ & 1  & 10 & $\bigcirc$\\ \hline
 L59I0d5  & 5.9 & --          & 5  & 10 & X\\
 L59I1d5  & 5.9 & $\times 1$  & 5  & 10 & X\\
 L59I3d5  & 5.9 & $\times 3$  & 5  & 10 & $\bigcirc$\\
 L59I0d10 & 5.9 & --          & 10 & 10 & $\bigcirc$\\ \hline
 L59T15   & 5.9 & $\times 1$  & 1  & 15 & X\\
 L59T20   & 5.9 & $\times 1$  & 1  & 20 & X\\
 L59T25   & 5.9 & $\times 1$  & 1  & 25 & $\bigcirc$\\ \hline
 L58I0    & 5.8 & --          & 1  & 10 & X\\
 L58I1    & 5.8 & $\times 1$  & 1  & 10 & X\\
 L58I10   & 5.8 & $\times 10$ & 1  & 10 & $\bigcirc$\\
 L58I30   & 5.8 & $\times 30$ & 1  & 10 & X\\  \hline
 L55I0    & 5.5 & --          & 1  & 10 & X\\
 L55I1    & 5.5 & $\times 1$  & 1  & 10 & X\\
 L55I10   & 5.5 & $\times 10$ & 1  & 10 & X\\
 L55I30   & 5.5 & $\times 30$ & 1  & 10 & X\\
\enddata
\tablecomments{%
$L_{\nu_{\rm e}}$ represents the luminosity of electron-type neutrino.
For $Q_{\rm inel}$, only the multiplicative factor is given.
$\delta v_{r}/v_{r}^{1D}$ denotes the initial relative amplitude
of velocity perturbation in percentage.
$T_{\nu_{\mu, \tau}}$ is the temperature of mu and tau neutrinos.
The ``successful shock revival'' is defined as a continuous increase
of the shock radius by $\sim$500~ms.
}
\end{deluxetable}

\clearpage

\clearpage

\clearpage

\clearpage

\clearpage

\clearpage

\end{document}